\documentclass[11pt,a4paper]{scrartcl}

\usepackage{ILD}

\usepackage[symbol]{footmisc}
\usepackage{feynmf}
\usepackage{textpos}
\usepackage{array,multirow,graphicx}
\usepackage{graphicx}
\usepackage{cleveref} 
\usepackage{tikz}
\usepackage[compat=1.1.0]{tikz-feynman}
\usetikzlibrary{arrows,shapes,positioning}
\usetikzlibrary{decorations.markings}
\usetikzlibrary{decorations.pathmorphing}
\usetikzlibrary{decorations.pathreplacing}
\usetikzlibrary{patterns}
\usetikzlibrary{plotmarks}
\usetikzlibrary{shadows}
\usepackage{contour}
\usepackage[style=numeric-comp,sorting=none]{biblatex}

\title{Conceptual aspects for the improvement of the reconstruction of $b$- and $c$-jets at $e^{+}e^{-}$ Higgs Factories with ParticleFlow detectors}

\ildproc{phys}{2021}{005}

\date{\today}

\addauthor{Yasser Radkhorrami}{\institute{1}\institute{2}}
\addauthor{Jenny List}{\institute{1}}

\addinstitute{1}{Deutsches Elektronen-Synchrotron (DESY), Hamburg}
\addinstitute{2}{Universit\"at Hamburg, Hamburg}

\abstract{The Higgs boson decay modes to $b$ and $c$ quarks are crucial for many Higgs precision measurements. The presence of semileptonic decays in the jets originating from $b$ and $c$ quarks causes missing energy due to the undetectable neutrinos. A correction for the missing neutrino momenta can be derived from the kinematics of the decay up to a two-fold ambiguity. The correct solution can be identified by a kinematic fit, which exploits the well-known initial state at an $e^{+}e^{-}$ collider by adjusting the measured quantities within their uncertainties to fulfill the kinematic constraints. The ParticleFlow concept, based on the reconstruction of individual particles in a jet allows understanding the individual jet-level uncertainties at an unprecedented level. The modeling of the jet uncertainties and the resulting fit performance will be discussed for the example of the ILD detector. Applied to $H\rightarrow b\bar{b}/c\bar{c}$ events, the combination of the neutrino correction with the kinematic fit improves the Higgs mass reconstruction significantly, both in terms of resolution and peak position.}

\titlecomment{Based on the talk "Reconstruction of $b$- and $c$-jets at $e^{+}e^{-}$ Higgs Factories with ParticleFlow detectors" presented at the International Workshop on Future Linear Colliders (LCWS2021), Geneva, Switzerland, 15-18 March 2021, PD3-40}

\graphicspath{ {./logos/}{./figures/} }

\begin{document}

\titlepage

\section{Introduction}\label{SEC:Intro}
Precision measurements of the 125-GeV Higgs boson are at the core of the physics program of future electron-positron colliders. Thereby, decays of the Higgs boson into heavy flavour jets play important roles at lepton colliders: decays into $b$ quarks because they are expected to be the most frequently occurring ones \cite{PDG}, decays into $c$ quarks because they are extremely challenging to measure at hadron colliders \cite{group2013handbook}. Several important measurements would thus profit significantly from further improvements of heavy flavour-jet reconstruction and identification.\\
Detector concepts of most Higgs factories are based on the Particle Flow paradigm \cite{Thomson_2009}. The International Large Detector (ILD) as one of the existing detector concepts of which detailed designs exist, is a highly granular detector optimized for particle flow algorithm \cite{theildconceptgroup2010international}. For ILD, the difference between $uds$ and $c$- or $b$-jets have been studied recently for the first time \cite[fig. 8.3d]{theildcollaboration2020international}. It has also been shown that in terms of the jet energy resolution, it's only the neutrinos from semi-leptonic decays that cause different behavior of $c$- and $b$-jets in comparison with jets originating from light $d$, $u$ and $s$ quarks. This comparison was studied using $Z\rightarrow q\bar{q}$ samples and $q\in [d,u,s,c,b]$ with five selected energies $E_{jet}$=(20, 45, 100, 175, 250) GeV. Table \ref{tab:SLDpercent} shows the percentage of the samples corresponding to certain number of semi-leptonic decays of $B$-hadrons ($N_{SLD}^{B}$) and $C$-hadrons ($N_{SLD}^{C}$). This shows that two thirds of samples will have at least one semi-leptonic decay.
\begin{table}[htbp]
	\centering
	\begin{tabular}{| c | c | c | c | c |}\cline{3-5}
		\multicolumn{2}{c|}{} & \multicolumn{3}{c|}{$N_{SLD}^{B}$} \\ \cline{3-5}
		\multicolumn{2}{c|}{} & 0 & 1 & 2 \\ \cline{1-5}
		\parbox[t]{4mm}{\multirow{3}{*}{\rotatebox[origin=c]{90}{$N_{SLD}^{C}$}}} & 0 & 34\% & 24\% & 4\% \\ \cline{2-5}
		 & 1 & 18\% & 12\% & 2\% \\ \cline{2-5}
		 & 2 & 3\% & 2\% & 0\% \\ \hline
	\end{tabular}
	\caption{Percentage of $Z\rightarrow b\bar{b}$ samples with certain number of semi-leptonic decays}
	\label{tab:SLDpercent}
\end{table}\\
The reconstruction of $b$-jets is studied on a MC sample of the process $e^{+}e^{-}\rightarrow ZH\rightarrow\mu\bar{\mu}b\bar{b}$ at a center-of-mass energy of 250 GeV, as produced for ILD's most recent MC production \cite{fullSM_ILC}. This sample includes initial state radiation (ISR) and the up-to-date 250 GeV beam energy spectrum, only the overlay from $\gamma\gamma\rightarrow$low $p_{T}$ hadron events has been omitted for the time being to focus on the $b$-jet reconstruction.
\begin{figure}[htbp] \centering
	\includegraphics[width=0.5\textwidth]{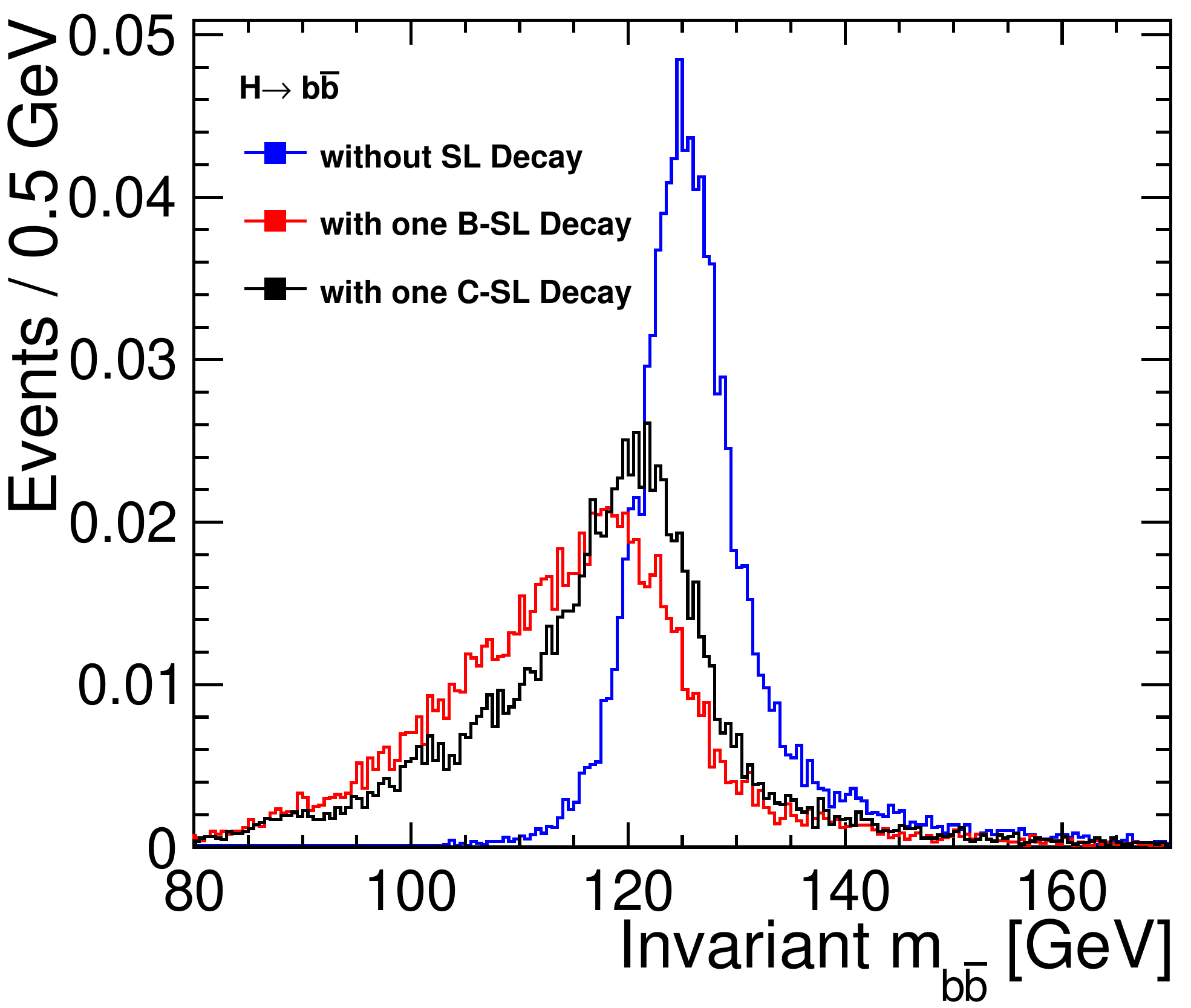}
	\caption{Mis-reconstruction of Higgs invariant mass due to semi-leptonic decay in $H\rightarrow b\bar{b}$ channel}
	\label{fig:InvMass_bb_nSLD}
\end{figure}\\
The presence of semi-leptonic decays causes di-jet invariant mass to be reconstructed at much lower values due to the missing neutrino energy. For instance in the $H\rightarrow b\bar{b}$ channel, the reconstructed Higgs invariant mass has worse peak position and resolution in presence of one semi-leptonic decay of $B$-hadron ($N_{SLD}^{B}=1$) or $C$-hadron ($N_{SLD}^{C}=1$) as illustrated in red and black histograms, respectively, in figure \ref{fig:InvMass_bb_nSLD}. This study will investigate to which extent the mass reconstruction can be improved by correcting for the missing neutrino momentum.
\section{Concept of the $\nu$-correction}\label{SEC:NuCorr}
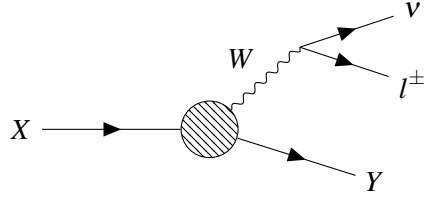
\begin{figure}[htbp] \centering
	\begin{tikzpicture}
		\begin{feynman}[scale=0.99]
			\vertex[blob] (m) at (0, 0) {\contour{white}{}};
			\vertex (H) at (-2.5, 0) {$X$};
			\vertex (X) at ( 2.2,-0.7) {$Y$};
			\vertex (W) at ( 1.2, 1.1 );
			\vertex (l) at ( 2.7, 0.6) {$l^{\pm}$};
			\vertex (nu) at ( 2.7, 1.6) {$\nu$};
			\diagram*
			{
				(H) -- [fermion] (m),
				(m) -- [fermion] (X),
				(m) -- [photon, edge label=$W$] (W),
				(W) -- [fermion] (l),
				(W) -- [fermion] (nu),
			};
			\draw[lightgray,ultra thin] (-3,-1) (3,1);
		\end{feynman}
	\end{tikzpicture}
	\caption{A schematic view of the semi-leptonic decay of $X$ ($B$/$C$ hadron) to an invisible neutrino and the visible $l^{\pm}$ and $Y$ decay products}
	\label{fig:XSLD}
\end{figure}

\begin{figure}[htbp] \centering
	\includegraphics[width=0.5\textwidth]{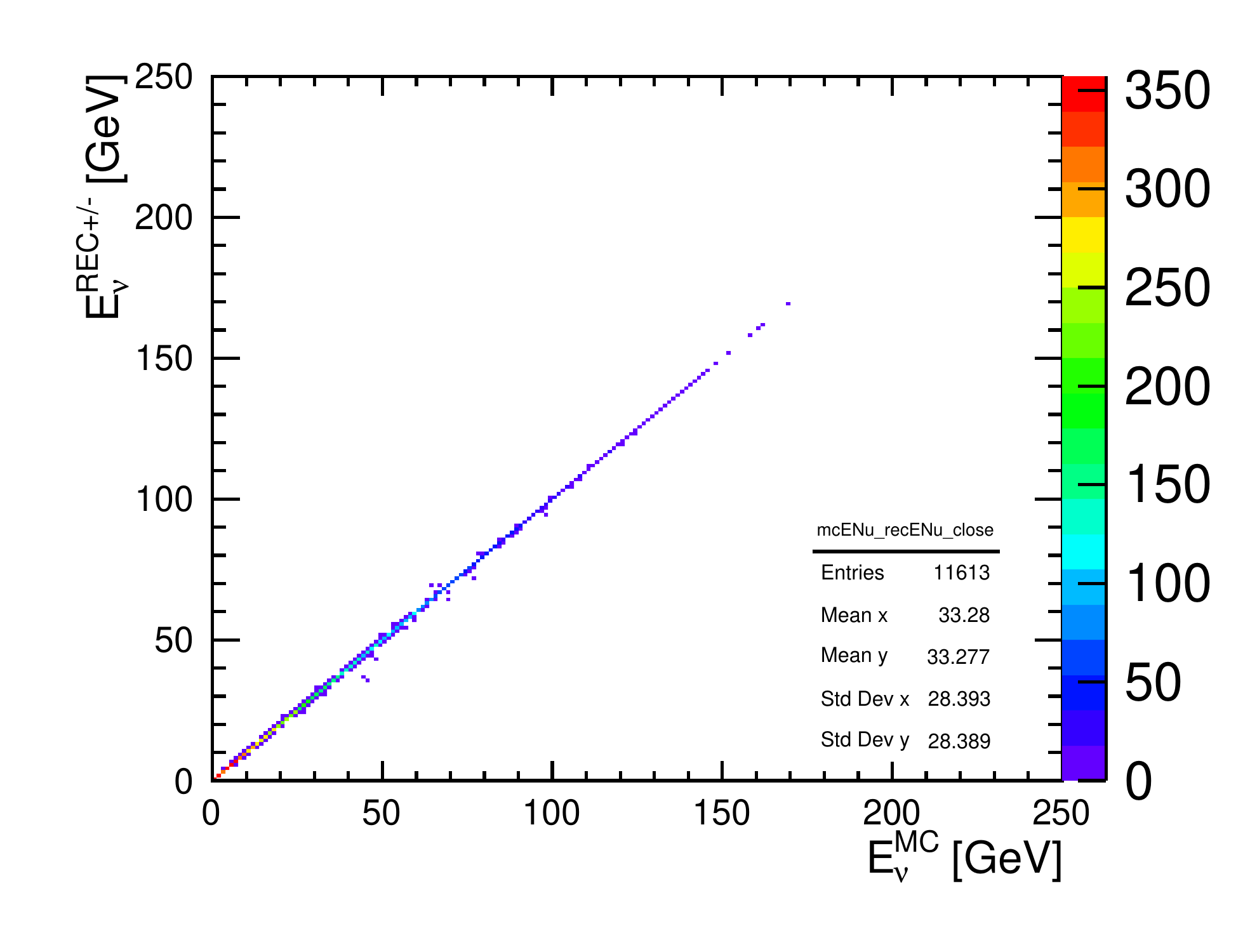}
	\caption{Inferred neutrino energy using the cheated information vs its MC truth value in $Z\rightarrow b\bar{b}$ samples with $E_{jet}=250$ GeV}
	\label{fig:recEnu_rapidity}
\end{figure}

The first step for correcting for the neutrino momentum is to identify $b$- or $c$-jets. This means flavour tagging is crucial for the $\nu$-correction. The next step is to find the semi-leptonic decay(s) in the jet. Since the neutrinos from semi-leptonic decays accompany the corresponding charged lepton(s), the presence of a charged lepton in a jet often implies a semi-leptonic decay. The high granularity of the ILD detector provides the opportunity to find and tag leptons in jets with high efficiency and purity \cite[sec. 8.1.4]{theildcollaboration2020international}. The neutrino momentum is then estimated from the kinematics of the semi-leptonic decay assuming the following information is available: The mass ($m_{X}$) and flight direction of the mother hadron and the 4-momentum ($\vec{p}_{vis}$,$E_{vis}$) of all visible decay products ($l^{\pm}$,$Y$) originating from the vertex at which the semi-leptonic decay occurs (Figure \ref{fig:XSLD}).
\\
The invariant mass of the mother hadron is given by
\begin{linenomath*}
	\begin{equation} \label{eq:mX_unprime}
		\begin{split}
			m_{X}^{2} & = E_{X}^{2}-|\vec{p}_{X}|^{2}\\
			& = (E_{vis}+E_{\nu})^{2}-|\vec{p}_{vis}+\vec{p}_{\nu}|^{2}
		\end{split}
	\end{equation}
\end{linenomath*}
Here, $m_{X}$ is Lorentz invariant and can be calculated in the most convenient system of reference, which clearly is the rest frame of $X$ (the \textit{primed} system from now on), where $\vec{p}_{vis}'=-\vec{p}_{\nu}'$, and simply
\begin{linenomath*}
	\begin{equation} \label{eq:mX_pvis}
		\begin{split}
			m_{X} & = E_{vis}' + E_{\nu}'\\
			& = E_{vis}' + |\vec{p}_{\nu}'|\\
			& = E_{vis}' + |\vec{p}_{vis}'|
		\end{split}
	\end{equation}
\end{linenomath*}
The neutrino is assumed to be massless and hence $E_{\nu}=|\vec{p}_{\nu}|$ in any frame. Taking the invariant mass of the visible decay products as $m_{vis}=\sqrt{E_{vis}'^{2}-|\vec{p}_{vis}'|^{2}}$, equation \ref{eq:mX_pvis} gives rise to
\begin{linenomath*}
	\begin{equation} \label{eq:Evis_prime}
			E_{vis}'=\frac{m_{X}^{2}+m_{vis}^{2}}{2m_{X}}
	\end{equation}
\end{linenomath*}
For the momentum of the visible decay products in the rest frame of $X$, one can write
\begin{linenomath*}
	\begin{equation} \label{eq:pvis}
		\begin{split}
			\vec{p}_{vis}'^{2} & = \vec{p}_{vis\parallel}'^{2}+\vec{p}_{vis\perp}'^{2}\\
			& = \vec{p}_{vis\parallel}'^{2}+\vec{p}_{vis\perp}^{2}
		\end{split}
	\end{equation}
\end{linenomath*}
where the parallel($\parallel$) and the perpendicular($\perp$) directions are taken with respect to the direction of flight of the mother hadron. The last step follows since the perpendicular component is not modified by the Lorentz transformation. Then
\begin{linenomath*}
	\begin{equation} \label{eq:pvis_prime}
		\begin{split}
			\vec{p}_{vis\parallel}' & = \pm \sqrt{\vec{p}_{vis}'^{2}-\vec{p}_{vis\perp}^{2}}\\
			& = \pm \sqrt{E_{vis}'^{2}-m_{vis}^{2}-\vec{p}_{vis\perp}^{2}}\\
			& = \pm \sqrt{(\frac{m_{X}^{2}+m_{vis}^{2}}{2m_{X}})^{2}-m_{vis}^{2}-\vec{p}_{vis\perp}^{2}}\\
			& = \pm \sqrt{(\frac{m_{X}^{2}-m_{vis}^{2}}{2m_{X}})^{2}-\vec{p}_{vis\perp}^{2}}
		\end{split}
	\end{equation}
\end{linenomath*}
The rapidity of the visible decays products of the semi-leptonic decay $\omega_{vis}$ in the lab frame is defined as
\begin{linenomath*}
	\begin{equation} \label{eq:def_rapidity}
		\omega_{vis}=\frac{1}{2}ln\frac{E_{vis}+\vec{p}_{vis\parallel}}{E_{vis}-\vec{p}_{vis\parallel}}
	\end{equation}
\end{linenomath*}
where $\vec{p}_{vis\parallel}$ is the component of visible momentum in the direction of flight of the mother hadron $X$. The rapidity behaves under Lorentz-transformations as the velocity does under Galileo-transformations. Thus
\begin{linenomath*}
	\begin{equation} \label{eq:add_rapidity}
		\omega_{X}=\omega_{vis}-\omega_{vis}'
	\end{equation}
\end{linenomath*}
where $\omega_{X}$ is the rapidity of mother hadron in the lab frame and $\omega_{vis}$ and $\omega_{vis}'$ are the rapidities of the visible decay products in the lab frame and in the rest frame of the mother hadron, respectively. Therefore, the primed values are related via the rapidity to the un-primed values which are measured in the lab frame. Taking the exponent of each side of equation \ref{eq:add_rapidity} gives
\begin{linenomath*}
	\begin{equation} \label{eq:Ehad}
		\begin{split}
			\frac{E_{X}+|\vec{p}_{X}|}{E_{X}-|\vec{p}_{X}|} & = \frac{E_{vis}+\vec{p}_{vis\parallel}}{E_{vis}-\vec{p}_{vis\parallel}}\frac{E_{vis}'-\vec{p}_{vis\parallel}'}{E_{vis}'+\vec{p}_{vis\parallel}'}\equiv C\\
			E_{X}+|\vec{p}_{X}| & = C(E_{X}-|\vec{p}_{X}|)\\
			(1-C)E_{X} & = -(1+C)|\vec{p}_{X}|\\
			& = -(1+C)\sqrt{E_{X}^{2}-m_{X}^{2}}\\
			E_{X} & = \frac{1+C}{2\sqrt{C}}m_{X}
		\end{split}
	\end{equation}
\end{linenomath*}
where
\begin{linenomath*}
	\begin{equation} \label{eq:C}
			\frac{1+C}{2\sqrt{C}} =\frac{E_{vis}E_{vis}'-\vec{p}_{vis\parallel}.\vec{p}_{vis\parallel}'}{m_{vis}^{2}+\vec{p}_{vis\perp}^{2}}
	\end{equation}
\end{linenomath*}
therefore
\begin{linenomath*}
	\begin{equation} \label{eq:EX}
		E_{X} = \frac{E_{vis}E_{vis}'-\vec{p}_{vis\parallel}.\vec{p}_{vis\parallel}'}{m_{vis}^{2}+\vec{p}_{vis\perp}^{2}}m_{X}
	\end{equation}
\end{linenomath*}
where $E_{vis}'$ and $|\vec{p}_{vis\parallel}'|$ are obtained from equations \ref{eq:Evis_prime} and \ref{eq:pvis_prime}, respectively, and the direction of $\vec{p}_{vis\parallel}'$ using the flight direction of the mother hadron. Finally, the energy of the neutrino in the lab frame is obtained by
\begin{linenomath*}
	\begin{equation} \label{eq:Enu}
		E_{\nu}=E_{X}-E_{vis}
	\end{equation}
\end{linenomath*}
where $E_{X}$ is given by equation \ref{eq:EX} and $E_{vis}$ is the measured energy of visible decay products. The last step is to calculate vectorial momentum of the neutrino by combining $\vec{p}_{\nu\perp}=-\vec{p}_{vis\perp}$ and equation \ref{eq:Enu}. Derived from equation \ref{eq:pvis_prime}, there will be a sign ambiguity in the obtained energy (and momentum) of the neutrino which is impossible to resolve without further information.\\
The flight direction of the mother hadron can be reconstructed from the line of flight between the primary and secondary (or secondary and tertiary) vertices. The mass of the mother hadron can be assumed to be the $B^{0}$ or $D^{0}$ mass, depending on the jet flavour and the rank of the vertex the lepton originates from. The $B^{0}$ and $D^{0}$ masses do not differ much from the $B^{\pm}$ and $D^{\pm}$ masses, covering together the majority of cases. The most daring assumption is the correct assignment of the visible particles to their vertex of origin. While with the precision of the vertex detectors this is expected to work very well for charged particles, the assignment of the neutrals needs to be studied in the future. For a first proof-of-principle of the neutrino correction, the flight direction of the mother hadron and the visible momentum at the decay vertex are cheated from MC truth. Figure \ref{fig:recEnu_rapidity} shows as a closure test the inferred neutrino energy using the cheated information vs its MC truth value in previously introduced $Z\rightarrow b\bar{b}$ samples with $E_{jet}=250$ GeV. Deciding between two solutions for neutrino energy is impossible without further information like event kinematics. Here, the solution giving the better match with the true energy is chosen.
\\
\section{Kinematic fitting}\label{SEC:KinFit}
\begin{figure}[htbp] \centering
	\begin{subfigure}{0.51\textwidth}
		\includegraphics[width=0.99\textwidth]{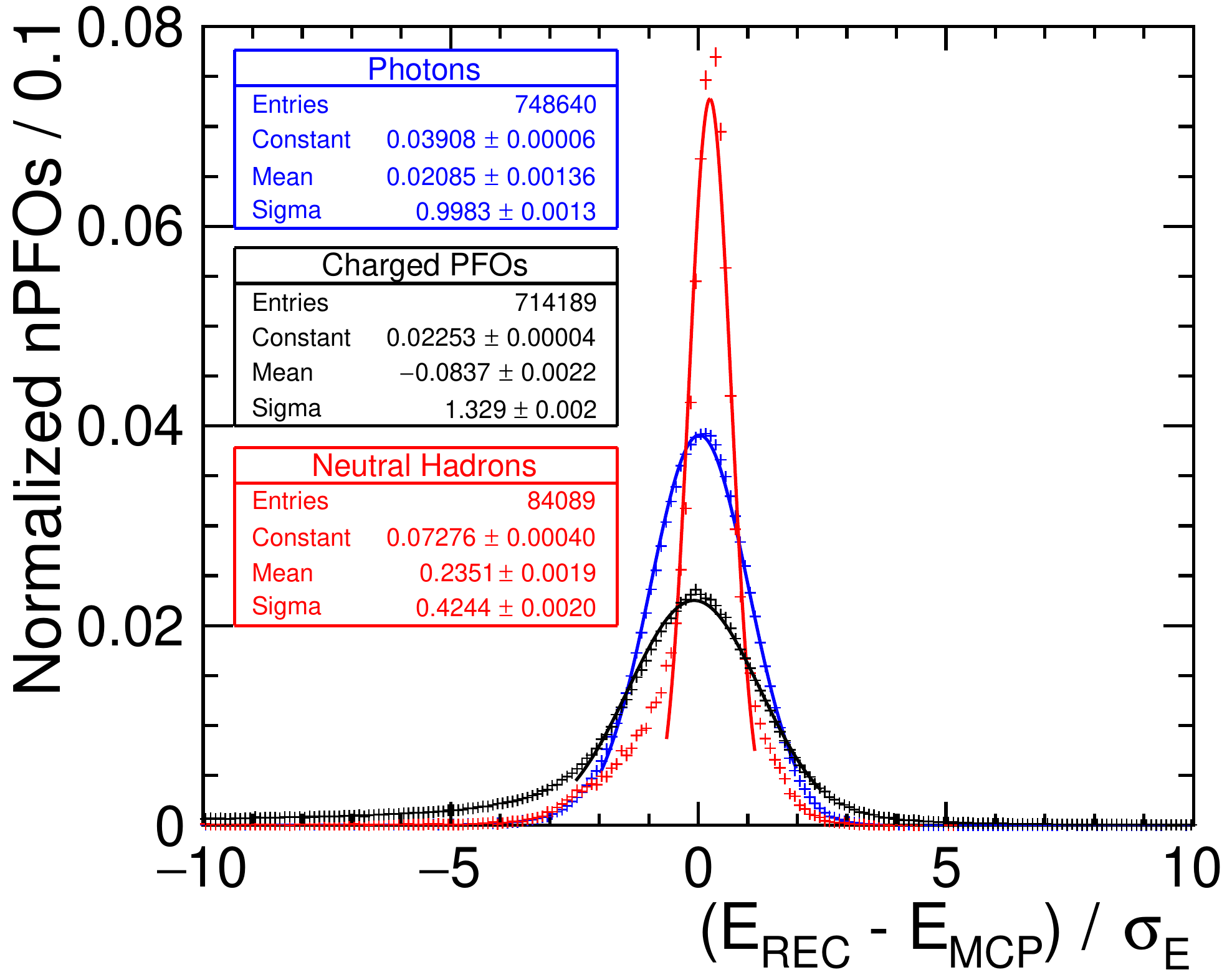}
		\caption{}
		\label{fig:sigmaE_pfo}
	\end{subfigure}
	\begin{subfigure}{0.48\textwidth}
		\includegraphics[width=0.99\textwidth]{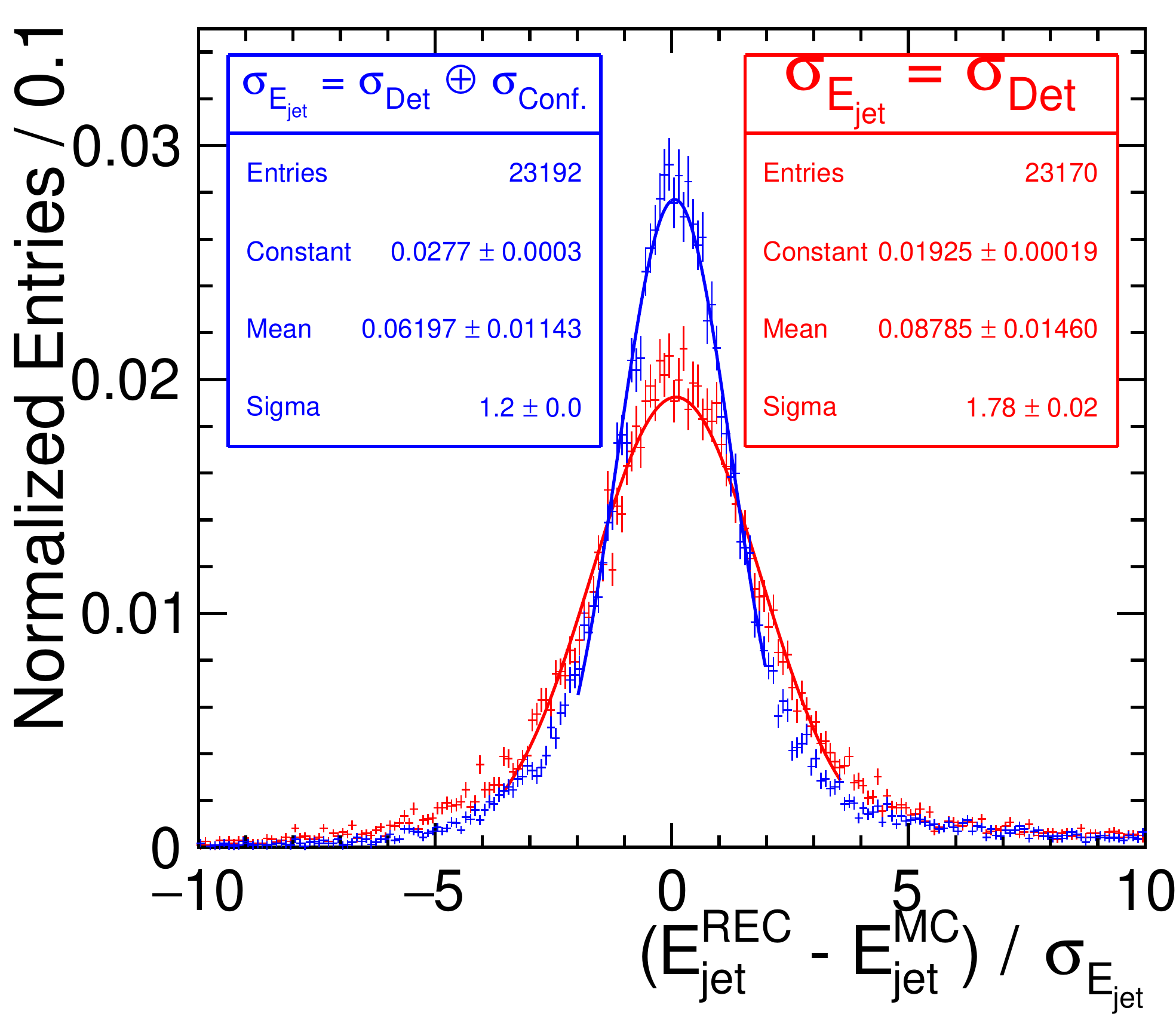}
		\caption{}
		\label{fig:sigmaE_jet}
	\end{subfigure}
	\caption{True energy ($E_{MCP}$) vs measured energy ($E_{REC}$) and energy error ($\sigma_{E}$). (a): at PFO-level, charged and neutral hadron PFOs need scaling factor of 0.75 nad 2.35, respectively, while photons are very well reconstructed. (b): at jet level, jet energy modeling is improved by adding uncertainty due to particle confusion ($\sigma_{conf}$).}
	\label{fig:sigmaE}
\end{figure}

\begin{figure}[htbp] \centering
	\begin{subfigure}{0.49\textwidth}
		\includegraphics[width=0.99\textwidth]{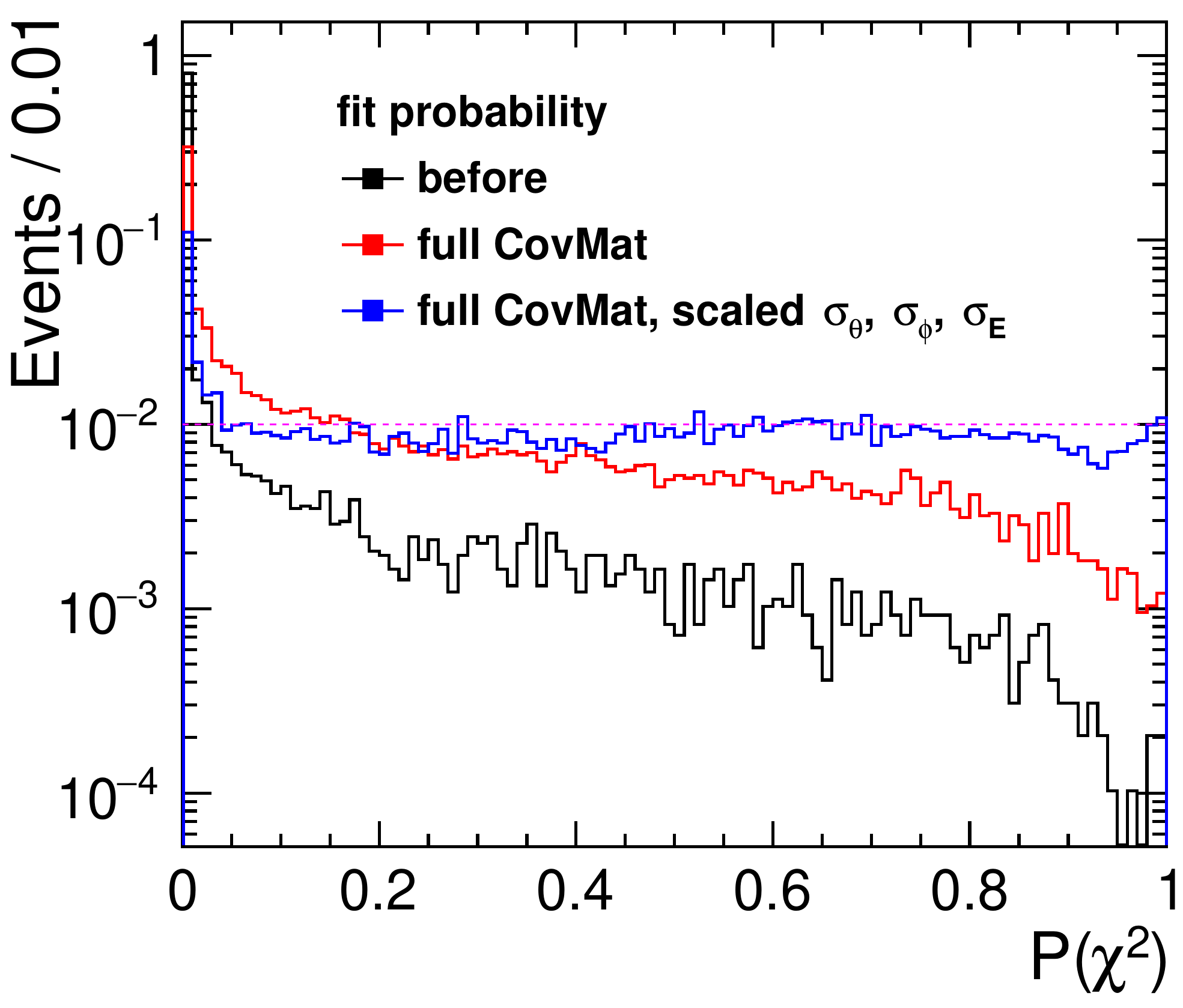}
		\caption{}
		\label{fig:FitProbA}
	\end{subfigure}
	\begin{subfigure}{0.49\textwidth}
		\includegraphics[width=0.99\textwidth]{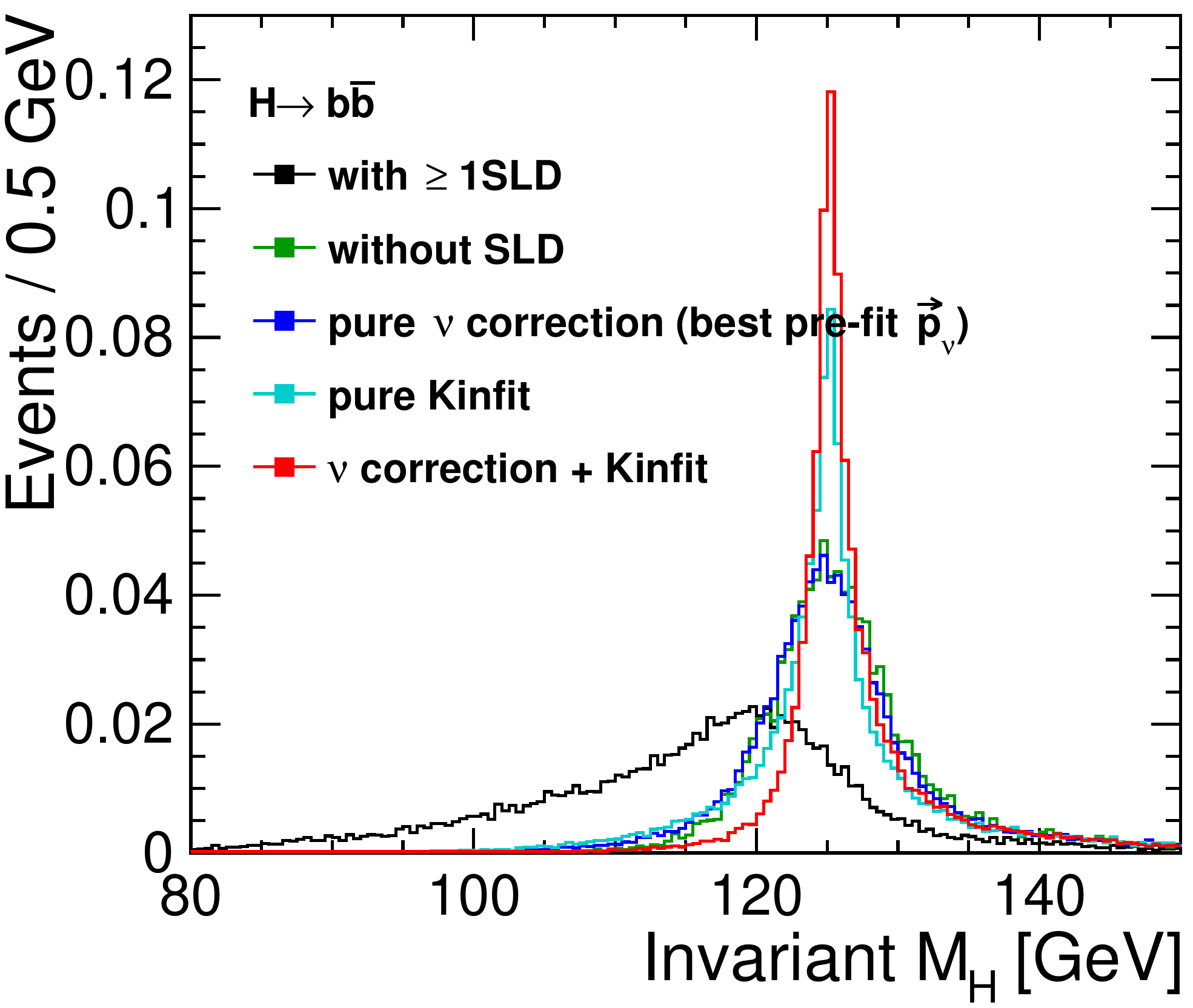}
		\caption{}
		\label{fig:FitProbB}
	\end{subfigure}
	\caption{(a): Propagating full covariance matrix of all PFOs to the jet level in addition to scaling obtained energy and angular uncertainties improve kinematic fit probability (only events without semi-leptonic decay). (b): Applying kinematic fit and neutrino correction gives drastic improvement on the Higgs mass reconstruction}
	\label{fig:FitProb}
\end{figure}

Thanks to clean collision environment at lepton colliders, the conservation laws can be used to improve the measurements beyond the detector resolution. A kinematic fitting technique is a mathematical tool that adjusts measured quantities within their uncertainties to fulfill certain constraints. For instance, the jet energy and di-jet invariant mass resolution is improved using kinematic fitting even in the presence of substantial ISR and Beamstrahlung \cite{list2009kinematic}. In our case, studying $ZH$ events, these constraints are energy and momentum conservation and the invariant mass of $Z$ boson as a soft constraint. The performance of the kinematic fitting is measured by fit probability and the pull distributions. This method is based on minimizing $\chi^{2}$:
\begin{linenomath*}
	\begin{equation} \label{eq:chi2}
		\chi^{2}(\pmb{a},\pmb{\xi},\pmb{f})=(\pmb{\eta}-\pmb{a})^{T}V^{-1}(\pmb{\eta}-\pmb{a})-2\lambda^{T}\pmb{f}(\pmb{a},\pmb{\xi})
	\end{equation}
\end{linenomath*}
where $\pmb{\eta}$ is the vector of the measured kinematic variables ($E_{jet}$, $\theta_{jet}$, ...), $\pmb{a}$ is the vector of fitted quantities, $\pmb{\xi}$ is the vector of the unmeasured kinematic variables, $V$ is the covariance matrix of the kinematic variables, $\lambda$ is Lagrange multipliers and $\pmb{f}(\pmb{a},\pmb{\xi})$ is the vector of constraints. An ideal kinematic fit will have a flat fit probability distribution and a Gaussian-shaped pull distribution centered at zero with unit width \cite{StatMethodsPhys}.\\
For being able to use kinematic fitting, one needs to have precise knowledge of the uncertainties of the measurements. For observables related to only one reconstructed object this is rather straight-forward: For instance the momentum of an isolated muon and its covariance matrix can in principle be derived from the covariance matrix of the track of the muon.\\
For jets, the situation is more complex. Still, in the particle flow approach, the uncertainties on the jet energy and angles can, in principle, be derived from the measurement uncertainties of the individual particle flow objects in the jets ($\sigma_{det}$) and estimates of the effects of the confusion in the particle flow ($\sigma_{conf}$), of mistakes in the jet clustering ($\sigma_{clus}$) and of the removal of overlay from $\gamma\gamma\rightarrow$low $p_{T}$ hadron events ($\sigma_{overlay}$). In case of our correction for semi-leptonic $b$- or $c$-decays, the estimate of the neutrino momentum will add an additional uncertainty ($\sigma_{\nu}$) \cite[sec. 7.2]{Ebrahimi:2017}.\\
In this study, we start by verifying the PFO-level uncertainties via the normalized residuals $\frac{x_{rec}-x_{true}}{\sigma_{x}}$. For charged PFOs, the uncertainties and covariances on the four-momentum are derived from the track fit. For neutral PFOs the corresponding information is obtained from the uncertainty on the energy and position measurement of the cluster in the calorimeter. Figure \ref{fig:sigmaE_pfo} shows as example the normalized residuals for the energy of charged PFOs, photon PFOs and neutral hadron PFOs. It turns out that the uncertainties on the PFO energy have to be scaled by factors of 0.75 and 2.35 for charged PFOs and neutral hadron PFOs, respectively, in order to obtain a width of about one. The energy and energy error of photons are measured very well.\\
The covariance matrices of the individual PFOs are then added up to form the most basic estimate for the jet covariance matrix. The corresponding normalized residual of the jet energy is shown in the red histogram in Figure \ref{fig:sigmaE_jet} for jets without semi-leptonic decays from events without overlay. The variance of the jet energies is clearly underestimated, leading to a width of about 1.8. When adding $\sigma_{conf}$ as derived in \cite{Ebrahimi:2017}, the width of the normalized residual improves considerably, as shown in the blue histogram in Fig \ref{fig:sigmaE_jet}. For the kinematic fit, the jet energy uncertainties are scaled up by 20\% motivated by the remaining width of the 1.2 instead of 1.0 for the normalized residual. Similarly, a scaling factor of 1.6 is derived for the uncertainties on the both jet polar and azimuthal angles. The jet energies and jet angles are assumed to be uncorrelated in the fit. Using the full 4-momentum covariance matrix could be a future improvement.
\\
The impact of this new jet error parametrisation is evaluated on the previously introduced $ZH\rightarrow\mu\bar{\mu}b\bar{b}$ sample. For this purpose, only events in which none of the $b$-jets contains a semi-leptonic decay are used. The resulting fit probability is compared in Figure \ref{fig:FitProbA} to the one obtained with the previous, simpler jet error parametrisation with fixed angular resolution ("before", black), and with the new parametrisation but excluding the extra jet-level scaling factors (red). The performance improvement in each step is striking, resulting in a nearly flat distribution of the fit probability with the scaling factors.\\
With these jet error parametrisations, we can use the kinematic fit to resolve the ambiguity in the neutrino momentum. Three hypotheses are tested for each charged lepton found in a jet: no additional neutrino, a neutrino with the forward solution (+ sign in eq \ref{eq:pvis_prime}) and with the backward solution (- sign in eq \ref{eq:pvis_prime}). Out of all the hypotheses tested, the one which yields the highest fit probability is chosen.\\
Figure \ref{fig:FitProbB} shows the effect on the reconstructed Higgs mass. The green histogram shows the invariant di-jet mass in the absence of semi-leptonic decays, corresponding to the blue histogram in Figure \ref{fig:InvMass_bb_nSLD}. The black histogram shows the invariant mass in all cases where there is at least one semi-leptonic decay involved, i.e. the sum of the red and black histograms in Figure \ref{fig:InvMass_bb_nSLD} plus all events with more than one semi-leptonic decay. The blue histogram shows the same events as in the black histogram, but with the neutrino correction applied. Here, the kinematic fit is just used to find the best solution, but the pre-fit values of the jet momenta (+ neutrino momenta) are used to calculate the invariant mass. The cyan and red histograms show the invariant mass obtained from the fitted jet momenta, without and with applying the neutrino correction, respectively. While the neutrino correction alone recovers a mass resolution very similar to the events without semi-leptonic decays, the mass reconstruction receives a further striking improvement when using the fitted jet momenta.
\section{Conclusions}\label{SEC:Concl}
The reconstruction of heavy flavour jets is very important for many Higgs measurements. A method to correct for the missing neutrino momentum from semi-leptonic decays based on detailed reconstruction of the decay kinematics has been proposed and tested at the conceptual level. The full information of the highly granular detectors optimised for particle flow reconstruction can be exploited to derive a jet-by-jet error parametrisation, which significantly improves the performance of kinematic fits. By combining the neutrino correction and the kinematic fitting, striking improvements of the Higgs mass reconstruction could be possible.\\
A full demonstration still requires to perform the neutrino correction based on reconstruction-level information (PFO momenta, vertexing etc) and a treatment of overlay from $\gamma\gamma\rightarrow$low $p_{T}$ hadron events.
\section{Acknowledgement}\label{SEC:Acknowl}
This work has benefited from computing services provided by the German National Analysis Facility (NAF)\cite{Haupt_2010}.
\printbibliography[title=References]
\end{document}